# Electrical Control of Dynamic Spin Splitting Induced by Exchange Interaction as Revealed by Time Resolved Kerr Rotation in a Degenerate Spin-Polarized Electron Gas

F. Zhang, H. Z. Zheng*, Y. Ji, J. Liu, G. R. Li

The State Key Laboratory for Superlattices and Microstructures, Institute of Semiconductors, Chinese Academy of Sciences, P. O. Box 912, Beijing 100083

PACS numbers: 72.25.Fe, 72.25.Dc, 78.47.+p, 71.70.Gm

The manipulation of spin degree of freedom have been demonstrated in spin polarized electron plasma in a heterostructure by using exchange-interaction induced dynamic spin splitting rather than the Rashba and Dresselhaus types, as revealed by time resolved Kerr rotation. The measured spin splitting increases from 0.256meV to 0.559meV as the bias varies from -0.3V to -0.6V. Both the sign switch of Kerr signal and the phase reversal of Larmor precessions have been observed with biases, which all fit into the framework of exchange-interaction-induced spin splitting. The electrical control of it may provide a new effective scheme for manipulating spin-selected transport in spin FET-like devices.

* To whom the correspondence should be addressed. E-mail: hzzheng@red.semi.ac.cn

The issue of most concern in spintronic devices is how to manipulate the spin degree of freedom in order to obtain superior functionality. The spin-orbit interactions both due to structure asymmetry (Rashba term) and due to bulk inversion asymmetry (Dresselhaus term) are going to induce spin splitting at nonzero in-plane wave vectors. They have extensively been explored to manipulate an assembly of spin polarized electrons in various kinds of prototype spin devices [1]. Because of in-plane momentum dependence, the spin-orbit interactions also bring a new type of precession-related spin dephasing events [2]. In the present work, instead of the Rashba and Dresselhaus types, it has been demonstrated that a dynamic spin splitting along the growth direction can be induced in heterostructures, when a population imbalance between two electron spin bands is created by circularly polarized excitation. That is because the single particle energy of an electron will be renormalized due to its exchange interaction with other electrons in interaction electron gas [3]. While such renormalization makes the single particle energy become smaller and depends on the population, the difference of the renormalized energy between the majority and minority spin bands, e.g., owing to an imbalance between their populations, gives rise to the observed dynamic spin splitting. The electrical control of it has been proved to provide a new effective scheme for manipulating spin-selected transport in spin FET-like devices.

Following the simplest physical picture, magneto-optical Faraday and Kerr effects appear as a result of dynamic bleaching observed in a time-resolved pump and probe measurement configuration. A population imbalance between majority and minority spin bands, created by the circularly polarized pump pulse, makes the optical coefficient of GaAs medium different, as probed with the opposite optical helicities ($\sigma^+$, $\sigma^-$). From the relevant theory [4], the KR angle in the reflection configuration may be expressed as

$$\theta_K = -\frac{\varepsilon'_{xy}}{n(n^2-1)}$$

$$= \frac{1}{n(n^2-1)} \frac{|M_\pm|^2_{vc}}{16\pi} \left(\frac{2m_r}{\hbar^2}\right)^{3/2} \left\{ \sqrt{\hbar\omega - E_g^+} \frac{1}{1+e^{\frac{\hbar\omega-E_g-\mu_F^+}{kT}}} - \sqrt{\hbar\omega - E_g^-} \frac{1}{1+e^{\frac{\hbar\omega-E_g-\mu_F^-}{kT}}} \right\} \quad (1),$$

if (1) the squared dipolar transition matrix elements, $|M_\pm|^2_{vc}$, for $\sigma^+$ and $\sigma^-$ helicities are approximately the same near the Γ point of non-magnetic semiconductors, (2) the effective mass approximation can be justified, and (3) the quasi-equilibrium condition for both spin-up and spin-down electrons is satisfied so that the original density functions in the conduction band, $\rho_{c+}$ and $\rho_{c-}$, can be replaced by the respective quasi-Fermi distribution functions with the quasi chemical potentials of $\mu_F^+$ and $\mu_F^-$ (which are measured from the respective bottoms of the majority and minority spin bands). Here, $m_r^{-1} = m_e^{*-1} + m_{hh}^{*-1}$; n is the refraction index of the medium; $E_g^+$ and $E_g^-$ are the effective band gaps for the split majority and minority spin bands. The other quantities are defined as usual. The above expression clearly describes the important features of KR at $E_g^+ = E_g^-$: (1) the magnitude of $\theta_k$ is proportional to $\sqrt{\hbar\omega - E_g}$, obeying the general rule for optical interband absorption, and (2) it increases with the population imbalance between the majority and minority spin bands. From Eq. (1), it also follows that the variation of time-resolved Kerr rotation (TRKR) with $\hbar\omega$, especially in the sign, could be employed to verify the dynamic band renormalization and the appearance of spin splitting between the majority and minority spin bands. On the other hand, in the framework of the mean-field theory, the exchange self energy is a negative term, and determined by $-\sum_q V_{|k-q|} f_q$, where $V_{|k-q|}$ is the Coulomb interaction between electrons, and $f_q$ is the Fermi distribution function [3]. When the dynamic population in two spin bands is imbalanced, the self energies for the majority and minority spin bands should also be different, leading to a dynamic spin-splitting given by $-\sum_q V_{|k-q|}(f_q^+ - f_q^-)$ [5]. If $f_q^+ > f_q^-$, the spin-down states, excited by $\sigma^+$ polarized pump pulses, are in the majority band, and their energy is down-shifted more than that of the spin minority band (retaining spin-up states). Hereafter, the designation of the majority and minority spin bands by superscripts $+$ and $-$, respectively, is retained throughout this article. Such exchange-interaction induced spin splitting was also employed to explore its influence on the spin dephasing process in a two-dimensional electron system (2DES) first theoretically [5] and then experimentally [6], as it acts an effective magnetic field along the growth direction.

Based on the idea above, the sample structure in investigation was designed as a single-barrier tunneling diode, grown by molecular beam epitaxy (MBE) with a thick intrinsic GaAs as the absorption layer. The layer structures were grown in the sequence: 250nm-thick Si-doped GaAs buffer layer (n= $1\times 10^{18}$cm$^{-3}$) on n$^+$-GaAs (100) substrate, 200nm-thick intrinsic GaAs, 5nm-thick intrinsic AlAs barrier, 500nm-thick intrinsic GaAs as the absorption region(labeled as W), 30nm-thick intrinsic Al$_{0.45}$Ga$_{0.55}$As and finally covered by a 100nm-thick Si-doped GaAs (n= $1\times 10^{18}$cm$^{-3}$, labeled as L). A 500 $\times$ 500 μm$^2$ optical window was defined by using photolithography and wet etching. An Ohmic contact was formed on the top surface by depositing and alloying Au/Ge/Ni metallic multilayers, the backside contact was formed in the same way on the n$^+$- GaAs (100) substrate. The sample was mounted inside a magneto-optical cryostat, the temperature of which varied from 1.5 to 300K and the magnet field of which scanned up to 10T.

Optical measurements were performed in a pump and probe configuration (see the inset to Fig. 1), using a mode-locked Ti:sapphire laser with a pulse duration of approximately 3ps and a repetition rate of 76 MHz, in the photon wavelength range from 700nm to 1000nm. The pump and probe beams were shined on the sample at an incidence angle of 0° and 5° (with respect to the sample normal) with a spot diameter of about 0.2 mm and the intensities of 1mw and 0.1mw respectively. A negative bias was applied on the top electrode with respect to the back side contact. An electronic feedback circuit programmed and maintained the relative delay time between pump and probe pulses with a temporal resolution of 1.7ps. In the measurement of scanning wavelength the accuracy of the wavelength is about 0.26A.

In Fig. 1, the KR was measured at a fixed probe delay time of 100ps under the biases of 0V, -0.3V and -0.6V by scanning the wavelengths of both the pump and probe beams simultaneously. As the wavelength starts to scan from 819nm, the KR undergoes a first oscillatory change before 816.7nm. Their overall behavior is roughly the same for 0V, -0.3V, -0.6V, while the amplitude increases slightly with increasing the bias. Since the wavelength range is still below the GaAs band gap, this oscillatory change naturally reflects the lifting of spin degeneracy of the excitons. Such spin splitting has previously been observed in polarized two dimensional exciton gas, and attributed to the repulsive interaction between excitons due to the Pauli exclusion principle [7]. The small kinks, appearing in the range from 817nm to 816.6nm, presumably arise from the spin-split excited levels of the excitons. It should, however, be emphasized that such repulsive interaction applies only for the two-level system with no energy-dispersion, like excitons. After the photon energy becomes larger than the band gap (at ~816.6nm for -0.3V, -0.6V, and ~816.7nm for 0V), the KR signals at three biases all reach their own positive maxima, indicating the dominant contribution from the polarized free electron gas in the conduction band. The positive sign is well in accordance with the theoretical expression (1). The most striking feature is that the sign reversal of the KR takes place under the biases of -0.3V and -0.6V, after the wavelength goes shorter than 816.1nm and 815.5nm, respectively. Meanwhile, the KR under zero bias remains positive over the whole scan range of the wavelengths. Judging from Eq. (1), one has to invoke the fact that only when a spin splitting appears between the majority and minority spin bands with large enough magnitude of $\Delta E_g = E_g^- - E_g^+$, then quasi Fermi level $E_F^-$ can be lifted above $E_F^+$ ($E_F^+$ and $E_F^-$ are measured from the top of the valence band, and have the relation of $E_F^\pm = E_g^\pm + \mu_F^\pm$). As a result, the sign of KR can be reversed on the short wavelength side. For clarity, the conduction bands modified by the exchange interaction are schematically shown in the inset to Fig. 2 for the case of 0.6V only for the sake of illustration. The conduction band edge for $E_g^+$ should correspond to the cross point at 816.6nm; $E_g^-$ is assigned to the wavelength of 816.3nm at the positive peak position; the cross point at 815.5nm roughly denotes the position of $E_F^+$. As the wavelength scans from 816.6nm to 816.3nm, only the first term in the parenthesis of Eq. (1) contributes to the KR in proportion to $\sqrt{\hbar\omega - E_g^+}$. As soon as the wavelength becomes shorter than 816.3nm, the second term in the parenthesis appears to diminish the KR continuously. The probe beam starts to probe the Kerr response only from the minority spin band, as its wavelength becomes shorter than 815.5nm, leading to the sign reversal of KR. In contrast to the cases of -0.3V and -0.6V, the spin splitting induced by the exchange interaction is rather small at the zero bias, so that $E_F^-$ lies always below $E_F^+$. As a result, the KR always remains positive, and the contribution from the minority spin band only partially cancels that from the majority spin band, as indicated by a sudden drop around 816nm in Fig. 2 (a). Since the delay time of the probe pulse is fixed at 100ps, the accumulation of the spin polarized electrons has nearly been completed in the vicinity of the AlAs barrier by drifting away the bulky GaAs layer. Accordingly, the quasi-equilibrium filling in both of the quasi-2D spin bands is almost fulfilled, and the variation of KR with the wavelength approximately reflects the phase-space filling in both spin bands and the spin splitting between them for the case where the photo-excitation is close to the band gap.

   To further verify the above point, the temporal transients of the KR were measured at fixed wavelength of 815nm under various biases in both the absence and presence of the magnetic field of 2 Teslas in the Voigt configuration. As shown in Fig. 3(a), the KR under zero bias shows a positive temporal response, primarily coming from the contribution of the 500nm-thick bulky GaAs layer. A small bias of -0.1V depresses the positive Kerr signal significantly. That may be ascribed to the fact that the number of spin polarized electrons, excited in the bulky GaAs layer, decreases due to fast drifting off in the electric field. Therefore, KR is substantially suppressed due to the reduction in the filled electron number at the energy probed by 815nm-photons. This may be visualized by comparing Fig. 2 (b) to Fig. 2 (a). It is intriguing that the sign of the KR is reversed by increasing the bias to -0.2V and -0.3V. In addition, a building-up process in the initial delay time of about 100ps also appears especially at -0.3V, indicating that the accumulation of spin polarized electrons towards the vicinity of AlAs barrier develops under the electric field. Because the Kerr response from the 500nm-thick bulky layer always remains positive as seen from Fig. 1, one may reasonably attribute the negative Kerr response at -0.2V and -0.3V to the contribution from the quasi-two-dimensional electron gas (Q-2DEG), as depicted by Fig. 2 (c). In the framework of the above physical picture, the continuously increased accumulation of the spin-polarized electrons, on one hand, lowers both the majority and minority spin bands with respect to the photon energy (at 815nm) due to the enhancement in the negative exchange self energy. On the other hand, it eventually makes the dynamic spin-splitting large enough to push $E_F^-$ above $E_F^+$. As long as the photon excitation at 815nm falls in the gap between $E_F^-$ and $E_F^+$, the reversal of the KR signal takes place as understood from Eq. (1) and illustrated by Fig. 2 (c).

   Similar variations in the phase and amplitude of the Larmor precession are also seen in the Fig. 3(b) under the different biases. It is found that the phase reversal of the Larmor precession takes place early at -0.1V, implying that the contribution from Q-2DEG is already admixed with that of the bulky layer. Both the phase and the amplitude of the Larmor precession change with the bias in accordance with the sign and the amplitude of Kerr signal at zero magnetic field.

The abnormal behavior within the initial 40ps may be attributed to the interference effect because of the drift of the excited spin-polarized electrons towards the AlAs barrier, observed recently by Salis [8]. Since it occurs quite early, it does not affect what concerns us here. Especially, if we trace back to Fig.1, where TRKR is measured in a scanning of wavelength at a fixed delay time(equivalently at a fixed distance of d), the phase factor $\phi = 4\pi n d / \lambda$, which determines the amplitude and phase of the interference, is impossible to vary dramatically, leading a KR sign reversal as the wavelength changes only by less than 0.1nm.

Although we attribute the phase reversal of the Larmor precession at -0.1V -0.2V and -0.3V to the dynamic spin splitting appearing between two spin bands of electrons, we still need to discuss some other possible mechanisms. First, it is well known that the dynamic nuclear polarization (DNP) could be induced along the pump beam by polarized optical excitation due to hyperfine exchange interaction between polarized electron spins and nuclear spins. The spin splitting in the conduction band by DNP should display significant difference, when the measurement is performed in the absence and in the presence of magnetic fields, because the Larmor precession in the latter case dephases the electron spin polarization transverse to the applied magnetic field substantially. A time-averaged spin $\overline{S_\perp}$ and its induced effective field will be smaller than that at zero field by three order of magnitudes [9]. One can never expect that the sign change of the KR in Fig. 3 (a) takes place synchronously with the phase reversal of the Larmor precession in Fig. 3 (b). Moreover, our observation persists until a temperature of 125K, at which DNP should wash away completely. Second, The spin-orbit interactions due to structure and bulk inversion asymmetries (Rashba and Dresselhaus effects), as a well established mechanism, can induce spin splitting at no zero in-plane wave vectors. From the Hamiltonian forms of Rashba and Dresselhaus spin-orbit interaction [2]( within k linear approximation): $H_{SIA} = \alpha(p_x\sigma_y - p_y\sigma_x)/\hbar$ and $H_{BIA} = \beta(p_x\sigma_x - p_y\sigma_y)/\hbar$ ( **p** is the momentum of the electron, and $\sigma$ the vector of Pauli matrices), it follows that their wave vector dependence makes the spin orientations are anti-parallel to each other at two opposite k vectors of same magnitude in ($k_x$, $k_y$) plane. No net spin polarization along any direction in the plane would be expected. Even if the spin polarization in the plane were present due to the presence of spin splitting induced by the spin-orbit interaction of Rashba and the Dresselhaus types, thus induced spin polarization would still have been irrelevant to the present observation. A linearly polarized probe beam, incident perpendicularly to the sample, can not sense a spin polarization transverse to it. Actually, we have carefully checked if any residue Larmor oscillation caused by the in-plane Rashba or Dresselhaus effective field could be observed in the zero magnetic field, as previously seen in [10]. But, nothing is found in Fig. 3 (a).

In what follows we make a rough estimation for the spin polarization degree and carrier accumulation required for our observation. Again, in the effective mass approximation, the dynamic spin-splitting will be estimated by considering only the statically screened 3D Coulomb potential in each spin band. By defining, $p = (n^+ - n^-)/(n^+ + n^-), n = n^+ + n^-$, one has $n^+ = (1+p)n/2$ and $n^- = (1-p)n/2$. Then, the screened Coulomb potential, screening vector and chemical potential are given as follows [3]:

$$V^\pm_{|k-q|} = 4\pi e^2 / \left[\varepsilon_0\left(\left|k^\pm - q\right|^2 + \kappa^2\right)\right], \quad \kappa = \sqrt{(6\pi e^2 n)/(\varepsilon_0 \mu_F)}, \quad n^\pm = (6\pi^2)^{-1}\left(2m^*\hbar^{-2}\mu_F^\pm\right)^{3/2}$$

where $\mu_F$ is calculated from the total carrier density n. For simplicity, the dynamic spin splitting $\Delta E_{exc}(k,T)$, given by $-\sum_q V_{|k-q|}\left(f_q^+ - f_q^-\right)$, is estimated at T=0K, then, it becomes

$$\Delta E_{exc}(k,0) = -\left\{\frac{1}{(2\pi)^2}\int_0^{k_F^+} q^2 dq \sin\theta d\theta \frac{4\pi e^2}{\varepsilon_0(k^2 + q^2 - 2kq\cos\theta + \kappa^2)} - \frac{1}{(2\pi)^2}\int_0^{k_F^-} q^2 dq \sin\theta d\theta \frac{4\pi e^2}{\varepsilon_0(q^2 + \kappa^2)}\right\}$$

Here, in order to match the measurement, the first integral for the majority spin band should be evaluated at k value determined from $k = \sqrt{2m\Delta E_{exp}(k,0)/\hbar^2}$ with $\Delta E_{exp}(k,0)$ being the measured spin splitting, and the second one at k=0. Let the following experimental parameters be adopted: the period of laser pulse T is 13ns, the absorption coefficient α is taken to be $10^4$ cm$^{-1}$, the light power P shone on the sample is about 1mW. The total density of carrier n is estimated from the excitation power, according to $n = (PT/\hbar\omega V)e^{-\alpha L}(1-e^{-\alpha W})$. By assuming that the degenerate electron gas is confined in a scale of W*=150nm at -0.3V, the volume V is taken to be $\pi D^2 W^*/4$ with D=0.2 mm. An itinerant calculation procedure gives a spin polarization degree p of 28% with a carrier density of $4.0\times 10^{15}$cm$^{-3}$. Next, the same value of p is used in the calculation for -0.6V. Now, the total carrier density accumulated in the vicinity of AlAs barrier will be determined in a self-consistent manner. It turns out that n= $1.3\times 10^{16}$cm$^{-3}$, which amounts to a confinement length of 46nm at -0.6V. The above estimation is rather rough, but, nevertheless, is instructive and reasonable. It illustrates us how an enhanced accumulation of spin polarized electron gas can affect the dynamic spin splitting. As a result, we are convinced that the present work shines some light on how one can manipulate the dynamic spin splitting under the circularly polarized photo-excitation by biasing the structure.

In conclusion, by applying bias, spin polarized electrons, excited by circularly polarized pump pulses in a 500nm-thick intrinsic GaAs layer, are accumulated into the vicinity of an AlAs barrier in a heterostructure. KR, measured as a function of wavelength at a fixed delay time of 100ps in a time-resolved pump and probe measurement configuration, reveals the appearance of the dynamic spin splitting between majority and minority spin bands. Both the spin splitting and the phase of the Larmor precession can electrically be controlled. For our case, the spin splitting induced by the exchange interaction increases from 0.256meV to 0.559meV as the bias varies from -0.3V to -0.6V. In contrast to the Rashba and Dresselhaus types, the exchange-interaction induced spin splitting is now along the growth direction, and it is more favorable for manipulating spin-selected transport in spin FET-like devices.

The authors would like to thank M. W. Wu for helpful discussion and Z. C. Niu for sample growth. This work was in part supported by National Basic Research Program of China No 2006CB932801and No2007CB924904, and also by Special Research Programs of Chinese Academy of Sciences.

Figure Captions

Fig. 1
KR was measured at 5K by simultaneously scanning wavelengths of both pump and probe pulses under different negative biases as the delay time of the probe pulse was fixed at 100ps.

Fig. 2
Band profiles under different negative biases with schematics of majority and minority spin bands plotted against to the photon energy of 815nm for bulk and quasi 2DEG, the physical content of them as explained in the text. The inset is the plot of majority and minority spin bands at -0.6V with the labels of the wavelengths at some of particular energy positions.

Fig. 3
(a) KR was measured at 5K under different negative biases of 0V, -0.1V, -0.2V, -0.3V at zero magnetic field as the wavelength is fixed at 815nm.
(b) Larmor precessions were displayed in KR under different negative biases of 0V, -0.1V, -0.2V, -0.3V at a magnetic field of 2 Teslas, as the wavelength is fixed at 815nm.

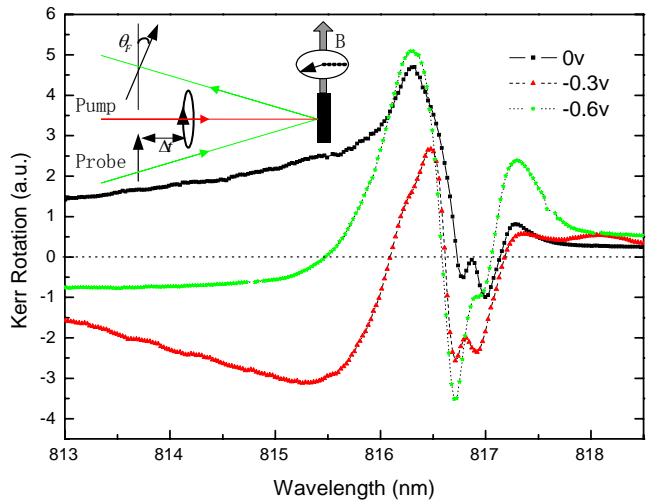

Fig. 1

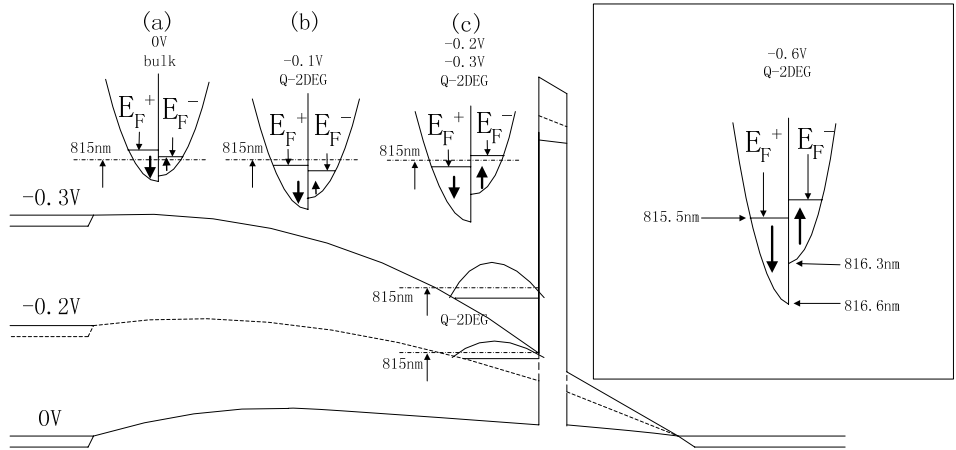

Fig. 2

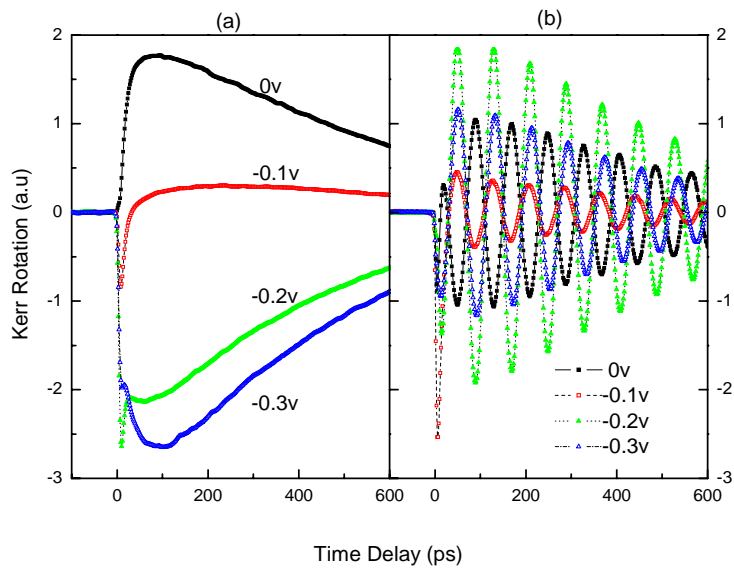

Fig. 3